\documentclass[a4paper, preprint,12pt,sort&compress]{elsarticle}

\usepackage{epsfig}

\usepackage[dvipdfmx]{color}

\usepackage{times} 
\usepackage{amssymb}
\usepackage{lineno}

\usepackage[utf8]{inputenc} 
\usepackage[T1]{fontenc} 
\usepackage{mathptmx} 

\journal{Journal of Magnetism and Magnetic Materials}

\begin{document}

\begin{frontmatter}

\title{Magnetic, thermal, and magnetocaloric properties of
the holmium trialuminide HoAl$_{3}$ with polytypic phases}

\author[1]{Takafumi D. Yamamoto\corref{cor1}}
\ead{YAMAMOTO.Takafumi@nims.go.jp}
\author[1,2]{Pedro Baptista de Castro}
\author[1]{Kensei Terashima}
\author[1]{Akiko T. Saito}
\author[1]{Hiroyuki Takeya}
\author[1,2]{Yoshihiko Takano}

\cortext[cor1]{Correpsonding author}

\address[1]{National Institute for Materials Science, Tsukuba, Ibaraki 305-0047, Japan}
\address[2]{University of Tsukuba, Tsukuba, Ibaraki 305-8577, Japan}

\begin{abstract}
We investigate the magnetic, thermal, and magnetocaloric properties of
the intermetallic HoAl$_{3}$ compounds with two different crystal structures.
The room-temperature equilibrium trigonal phase HoAl$_{3}$ undergoes
an antiferromagnetic (AFM) transition at
the N$\acute{\rm e}$el temperature $T^{\rm tri}_{\rm N} =$  9.8 K.
The AFM ordering of the compound is strong against the magnetic field,
so that inverse magnetocaloric effect (MCE) is found
even under a magnetic field change of 0--50 kOe.
The high-pressure cubic phase HoAl$_{3}$, prepared by a rapid-solidification process, is
antiferromagnetically ordered below $T^{\rm cub}_{\rm N} =$ 15 K.
Magnetic and specific heat measurements reveal that
this long-range AFM state becomes unstable as the temperature drops
and then it is replaced by a magnetic glassy state
below a spin freezing temperature $T_{\rm f} =$ 11 K.
The successive changes in magnetic state result in
complicated temperature and field dependence of the MCE at low temperatures.
\end{abstract}


\end{frontmatter}
\newpage
\section{Introduction}
Magnetocaloric effect is a magneto-thermodynamic phenomenon
in which a magnetic material absorbs/generates heat
when an applied magnetic field changes.
It is often characterized by the isothermal magnetic entropy change $\Delta S_{\rm M}$.
Since the discovery of the giant MCE of Gd$_{5}$Si$_{2}$Ge$_{2}$ in 1997
\cite{Pecharsky-PRL-1997},
much effort has been put into searching for magnetic materials with large MCEs,
the so-called magnetocaloric materials,
owing to their potential application to magnetic refrigeration near room temperature
\cite{Gschneidner-RPP-2005,Kitanovski-Springer-2015,Lyubina-JPD-2017}.
Recently, with the growing interest in magnetic refrigerators
for liquefaction of nitrogen, hydrogen, and helium,
magnetocaloric materials for low temperature applications have been
enthusiastically investigated for a wide range of materials,
from rare-earth based alloys
\cite{Li-SSC-2014,Matsumoto-JMMM-2017,Zhang-PhysicaB-2019,Zhang-JAC-2019,Pedro-NPG-2020,Li-JAC-2020}
to rare-earth based oxides
\cite{Omote-Cryogenics-2019,Li-ActMat-2020,Wu-CerInt-2021,Xu-MatTod-2021,Zhang-ActMat-2022}.

Binary holmium aluminide is one of a series of compounds
that has been widely studied in the context of magnetocaloric material exploration.
Among them, HoAl$_{2}$ is possibly the most well-known compound,
exhibiting a large $\Delta S_{\rm M}$ of $-$28.8 J kg$^{-1}$ K$^{-1}$
at the Curie temperature $T_{\rm C} \sim 30$ K
for a magnetic field change $\Delta H$ of 50 kOe
\cite{Hahimoto-ACEM-1986}.
These characteristics make this compound a promising candidate for
magnetic refrigerants for hydrogen liquefaction.
HoAl and Ho$_{3}$Al$_{2}$ are also reported as candidates
for low-temperature magnetic refrigerants
\cite{Yang-JAC-2014, Zhang-SSC-2012}.
The maximum value of $\Delta S_{\rm M}$ are
$-$22.5 J kg$^{-1}$ K$^{-1}$ at $T_{\rm C} =$ 20 K for HoAl
and $-$18.7 J kg$^{-1}$ K$^{-1}$ at $T_{\rm C} =$ 40 K for Ho$_{3}$Al$_{2}$
for $\Delta H =$ 50 kOe, respectively.
Besides, Ho$_{2}$Al is known to show multiple magnetic transitions at 40 K and 13 K,
with the maximum $\Delta S_{\rm M}$ of $-$6.9 J kg$^{-1}$ K$^{-1}$
around 30 K for $\Delta H =$ 50 kOe
\cite{Bhattacharyya-JPCM-2010}.
Curiously, as far as we know, only the trialuminide HoAl$_{3}$ has not yet been clarified
in terms of the magnetocaloric properties.

The crystal structure of \textit{R}Al$_{3}$ (\textit{R} $=$ Rare-earth) has been investigated in detail,
and it is known that it changes with
the temperature, pressure, and atomic number of the rare-earth elements
\cite{Vucht-JLCM-1965,Cannon-JLCM-1975}.
In contrast, little information is available on their magnetism,
only revealing that the room-temperature equilibrium phase \textit{R}Al$_{3}$ are
antiferromagnets with the N$\acute{\rm e}$el temperature $T_{\rm N}$ of 20 K or less
\cite{Buschow-HFM-1980}.
Meanwhile, it has been suggested that
the high-pressure phase \textit{R}Al$_{3}$ (\textit{R} $=$ Dy, Ho, and Er) are 
ferromagnets with MCEs comparable \textit{R}Al$_{2}$
\cite{Sahashi-IEEETrans-1987}.
These \textit{R}Al$_{3}$ systems in this work, however, are reported as
impurity phases in Al-rich \textit{R}Al$_{2}$ compounds,
so the physical properties of themselves still have room for verification.

\begin{figure}[t]
\centering
\includegraphics[width=150mm]{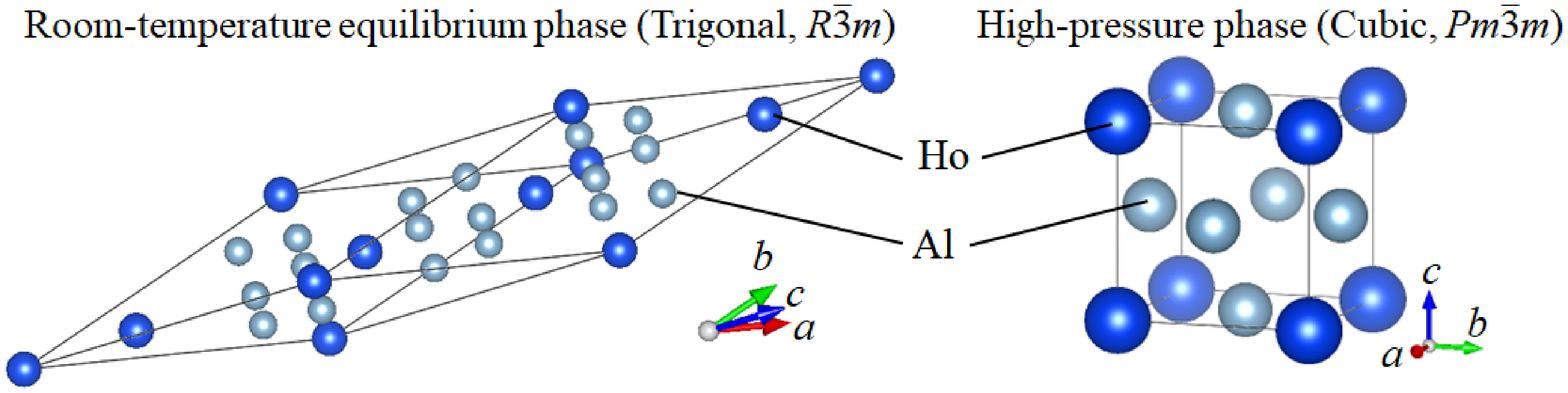}
\caption{(Color Online) Crystal structures of HoAl$_{3}$, drawn using VESTA \cite{Momma-VESTA-2011}.}
\label{fig:Structure}
\end{figure}

In this study, we comprehensively investigate on
the magnetic, thermal, and magnetocaloric properties of HoAl$_{3}$
for both the room-temperature equilibrium phase and the high-pressure phase.
As shown in Fig. \ref{fig:Structure},
the former has the HoAl$_{3}$-type trigonal structure
with the space group of \textit{R$\bar{3}$m},
and the latter has the Cu$_{3}$Au-type cubic structure
with the space group of \textit{Pm$\bar{3}$m}.
Here we attempt to synthesize the cubic phase HoAl$_{3}$
by a rapid-solidification process in the trigonal phase HoAl$_{3}$,
referring that high-pressure conditions have been simulated
in the rapidly solidified DyAl$_{3}$
\cite{Xu-APL-1991}.
The obtained samples are examined in terms of
the crystal structure, phase fraction, and physical properties.

\section{Material and Methods}
Starting material was prepared by arc-melting from Ho metal (3N) and Al metal (4N).
To avoid the appearance of HoAl$_{2}$
\cite{Meyer-JLCM-1966},
non-stoichiometric amounts of these reagents---Ho: Al $=$ 15 at.\%: 85 at.\%---were
melted and flipped several times in an arc-melting furnace under an argon atmosphere.
The rapid-solidification process was performed for the arc-melted alloy
by using another arc-melting furnace equipped with a copper hammer set.
A piece of the alloy 3--5 mm in size was placed on the copper hearth,
arc-melted again under an argon atmosphere,
and then rapidly quenched by hitting it with the hammer.
The splattered samples were collected for measurements.
Powder X-ray diffraction (XRD) measurements were performed
using a Rigaku MiniFlex600 diffractometer with Cu-$K\alpha$ radiation.
The XRD data were analyzed by a Rietveld refinement method
using the XERUS open-source package
\cite{Pedro-2022}.
The energy-dispersive X-ray spectroscopy (EDS) analysis was performed
to confirm element distribution of the samples
by using a JEOL JSM-6010LA scanning electron microscope (SEM) operated at 20 kV.
Magnetization ($M$) measurements were carried out by a Quantum Design SQUID magnetometer.
Specific heat ($C$) data were collected by a thermal relaxation method
with a Quantum Design PPMS's option.
The same sample was used in these physical property measurements.

\section{Results and discussion}
\subsection{Sample characterization}
\begin{figure}[t]
\centering
\includegraphics[width=120mm]{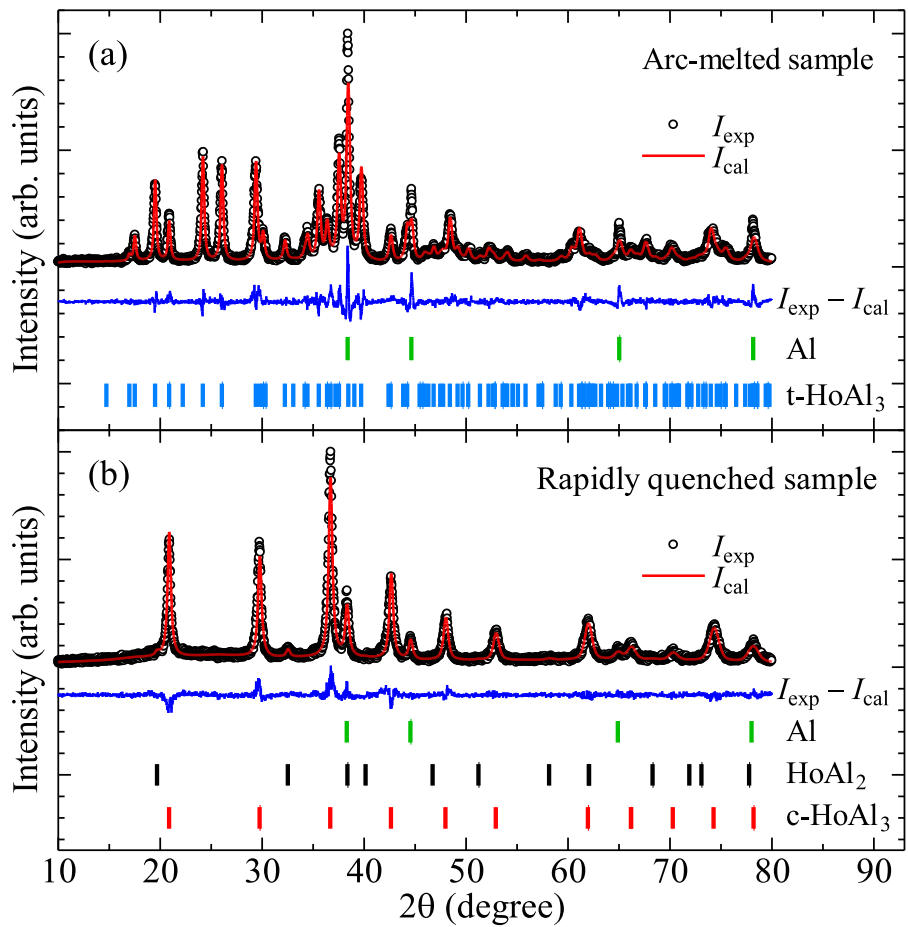}
\caption{(Color Online) Powder XRD patterns of the arc-melted HoAl$_{3}$ (a)
and the rapidly quenched HoAl$_{3}$ (b) along with Rietveld refinement.
The bars represent the Bragg peak positions of individual phases.
The t-HoAl$_{3}$ and the c-HoAl$_{3}$ indicate
trigonal phase HoAl$_{3}$ and cubic phase HoAl$_{3}$,
respectively.}
\label{fig:XRD}
\end{figure}

Figures \ref{fig:XRD}(a) and \ref{fig:XRD}(b) shows the powder XRD patterns of
the arc-melted HoAl$_{3}$ and the rapidly quenched HoAl$_{3}$.
The results demonstrate that the diffraction pattern is completely different
before and after the rapid-solidification process.
The main peaks can be assigned to the trigonal structure for the arc-melted sample
and to the cubic structure for the rapidly quenched sample.
Thus, it is confirmed that the rapid-solidification process causes
the transformation of the crystal structure of HoAl$_{3}$,
from the room-temperature equilibrium trigonal phase to the high-pressure cubic phase.
In this regard, the cubic phase HoAl$_{3}$ can be regarded as a metastable phase.
Besides, the change in crystal structure of HoAl$_{3}$ by the rapid quenching is highly reproducible.

The Rietveld refinement reveals that each sample contains impurity phases:
27 wt.\% of Al phase for both samples
and 3 wt.\% of HoAl$_{2}$ phase for the rapidly quenched sample.
The presence of a large amount of Al phase is
in consequence of the Al-rich nominal composition ratio.
The HoAl$_{2}$ phase was not found in the arc-melted sample by the Rietveld refinement,
implying that the weight fraction is less than 1\%.
Hereafter, we refer the arc-melted sample and the rapidly quenched one
to trigonal HoAl$_{3}$ and cubic HoAl$_{3}$, respectively.

\begin{figure}[t]
\centering
\includegraphics[width=120mm]{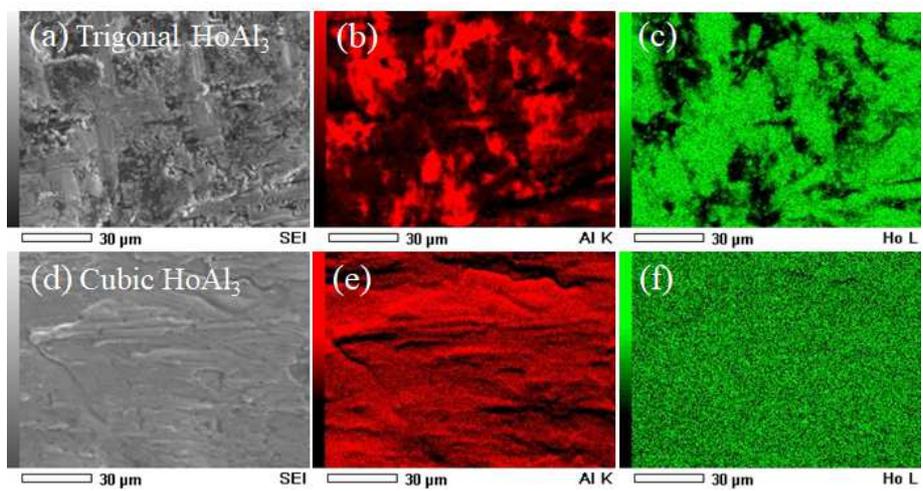}
\caption{(Color Online) (a)-(c) SEM secondary electron image (a) of the trigonal HoAl$_{3}$
and the corresponding EDS mapping images for Al (b) and Ho (c) elements.
(d)-(f) Those of the cubic HoAl$_{3}$.}
\label{fig:SEM}
\end{figure}

The SEM secondary electron images and the corresponding EDS mapping images
for Ho and Al elements are shown
in Figs. \ref{fig:SEM}(a)-(f) for each sample.
The trigonal HoAl$_{3}$ has Al-rich regions and Ho-rich regions.
The composition analysis at several positions reveals that
the former areas have 93--99 at.\% of Al element.
On the other hand, the composition ratio in the latter areas is evaluated to be
Ho: Al $=$ 27.5 at.\%: 72.5 at.\%,
roughly corresponding to HoAl$_{3}$.
Consequently, it is evident that the excess Al phase and the HoAl$_{3}$ phase are
well-separated each other in the trigonal HoAl$_{3}$.
Meanwhile, such phase separation is less visible in the cubic HoAl$_{3}$,
especially as can be seen from Fig. \ref{fig:SEM}(f).
The composition ratio approximately agrees with the nominal composition,
Ho: Al $=$ 15 at.\%: 85 at.\%,
over the entire region.
The excess Al phase and the HoAl$_{3}$ phase seem to be
homogeneously distributed in the cubic HoAl$_{3}$.

Here we will mention the influences of the impurity phases
on the thermodynamic properties of the sample.
The ferromagnetic HoAl$_{2}$ phase may contribute to
both the magnetization and the specific heat.
The non-magnetic Al phase would not contribute to the former at least in the low-temperature region,
but it may have a significant contribution to the latter.
The primary influence of the excess Al phase is
to largely reduce the phase fraction of HoAl$_{3}$ phase.
Accordingly, the observed values of the thermodynamic quantities are
lower than the real values for the studied compounds.
In this paper, the thermodynamic property data are given
in weight units using the total mass of the sample.
By combining these data with the weight fraction of the impurity phases,
one can estimate the real values of the thermodynamic quantities of HoAl$_{3}$ phase.

\subsection{Magnetic properties of trigonal HoAl$_{3}$}
\begin{figure}[t]
\centering
\includegraphics[width=100mm]{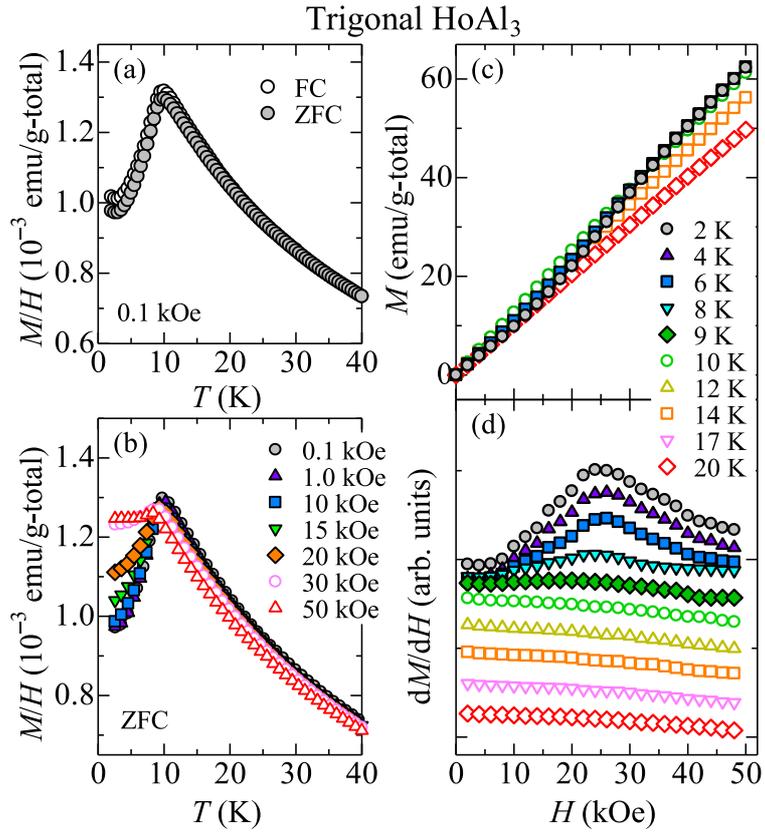}
\caption{(Color Online) Magnetic Properties of the trigonal HoAl$_{3}$.
(a) The temperature dependence of the magnetic susceptibility ($M/H$-$T$ curve)
at 0.1 kOe in ZFC and FC processes.
(b) $M/H$-$T$ curves under various magnetic fields in ZFC processes.
(c)-(d) Field dependence of the magnetization $M$ (c) and
the field derivative d$M$/d$H$ (d) at various temperatures.}
\label{fig:Mag-trigonal}
\end{figure}

Figure \ref{fig:Mag-trigonal}(a) shows the temperature dependence of
the magnetic susceptibility ($M/H$) of the trigonal HoAl$_{3}$ at 0.1 kOe
in zero field cooling (ZFC) and field cooling (FC) processes.
The sample was cooled down to 2 K in zero magnetic field before measuring the ZFC curve,
and then the magnetization data was taken at 0.1 kOe while heating up to 40 K.
Subsequently, the FC curve was measured while cooling under the same magnetic field.
A sharp cusp is observed at about 10 K,
being attributed to the occurrence of an AFM ordering reported in the literature
\cite{Buschow-HFM-1980}.
The N$\acute{\rm e}$el temperature $T^{\rm tri}_{\rm N}$, defined as the cusp temperature,
is evaluated to be 9.8 K.
This is consistent with the literature values.
As shown in Fig. \ref{fig:Mag-trigonal}(b),
the cusp is well-defined even in higher magnetic fields,
and the cusp temperature is slightly decreased down to about 8.3 K at 50 kOe.
These results suggest that the AFM ordering of the trigonal HoAl$_{3}$ is strong
against the magnetic field.

One finds that $M/H$ below the cusp temperature suddenly increases between 20 kOe and 30 kOe
and then becomes temperature-independent at 50 kOe.
For further investigation on this point,
we measured the field dependence of the magnetization.
Figure \ref{fig:Mag-trigonal}(c) shows the results at several temperatures.
The magnetization exhibits a linear field dependence at above 10 K,
whereas non-monotonical behavior is seen at even lower temperatures.
To make this clear, we show the field derivative d$M$/d$H$
at various temperatures in Fig. \ref{fig:Mag-trigonal}(d).
It is found that a broad peak appears around 25 kOe at below 9 K
and becomes pronounced with decreasing temperature.
These results imply a field-induced magnetic transition in the trigonal HoAl$_{3}$,
but the nature of which is to be further examined.
Nevertheless, at least the system seems not to be
in a (forced) ferromagnetic state even at 50 kOe
because the temperature dependence of $M/H$ at 50 kOe is different from
what is expected in such cases
\cite{Hu-APL-2008}.

We notice a small difference at below 15 K between ZFC and FC curves at 0.1 kOe.
One may consider that it is due to a tiny amount of the ferromagnetic impurity HoAl$_{2}$,
but this possibility can be ruled out.
According to the literature \cite{Khan-JAP-2013},
if this was the case, $M$/$H$-$T$ curves should exhibit an anomaly near
the Curie temperature of HoAl$_{2}$ ($\sim$ 30 K). 
The origin is unclear at the present stage.
Nonetheless, we will mention that no difference is found at above 1 kOe (not shown).

\subsection{Magnetic properties of cubic HoAl$_{3}$}
\begin{figure}[t]
\centering
\includegraphics[width=100mm]{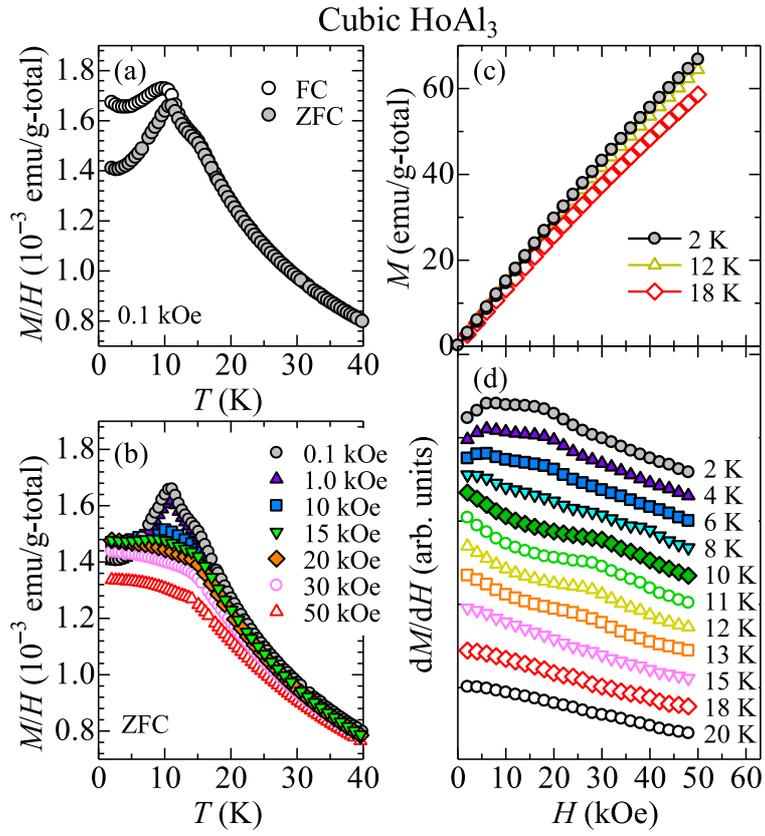}
\caption{(Color Online) Magnetic Properties of the cubic HoAl$_{3}$.
(a) $M/H$-$T$ curves at 0.1 kOe in ZFC and FC processes.
(b) $M/H$-$T$ curves under various magnetic field in ZFC processes.
(c)-(d) Field dependence of $M$ (c) and d$M$/d$H$ (d) at various temperatures.}
\label{fig:Mag-cubic}
\end{figure}

Figure \ref{fig:Mag-cubic}(a) shows the temperature dependence of $M/H$ of
the cubic HoAl$_{3}$ measured at 0.1 kOe in ZFC and FC processes.
Unlike the trigonal HoAl$_{3}$,
the $M$/$H$ of the cubic HoAl$_{3}$ exhibits two anomalies,
namely, a kink at 15 K and a cusp at 11 K.
As shown in Fig. \ref{fig:Mag-cubic}(b),
the kink is clearly observed even in higher magnetic fields,
and the kink temperature is only decreased by 0.9 K at 50 kOe.
These characteristics resemble to those of the cusp observed in the trigonal HoAl$_{3}$,
so it is expected that the kink at 15 K does originate from an AFM transition.
On the other hand, the cusp of the cubic HoAl$_{3}$ becomes obscured with increasing $H$
and disappears at above 20 kOe.
Moreover, a large difference between ZFC and FC curves is observed below the cusp temperature.
These results are quite different from those of the trigonal HoAl$_{3}$,
suggesting that the cusp of the cubic HoAl$_{3}$ at 11 K is not due to an AFM transition,
but rather can be ascribed to the occurrence of other magnetic states.
Note that no trace of HoAl$_{2}$ phase is found in $M$/$H$-$T$ curves near 30 K.

One can also find additional signs of a change in magnetic state of cubic HoAl$_{3}$
from the field dependence of the magnetization.
As shown in Fig. \ref{fig:Mag-cubic}(c),
the magnetization shows a monotonical field dependence at 2 K,
unlike the trigonal HoAl$_{3}$.
Furthermore, a broad peak of d$M$/d$H$ is observed around 30 kOe between 10 K and 13 K,
whereas it disappears at below 8 K
(see Fig. \ref{fig:Mag-cubic}(d)).
These results imply that the magnetic state of the cubic HoAl$_{3}$ is
no longer in an AFM state as the temperature decreases.

\subsection{Thermal properties}
\begin{figure}[t]
\centering
\includegraphics[width=100mm]{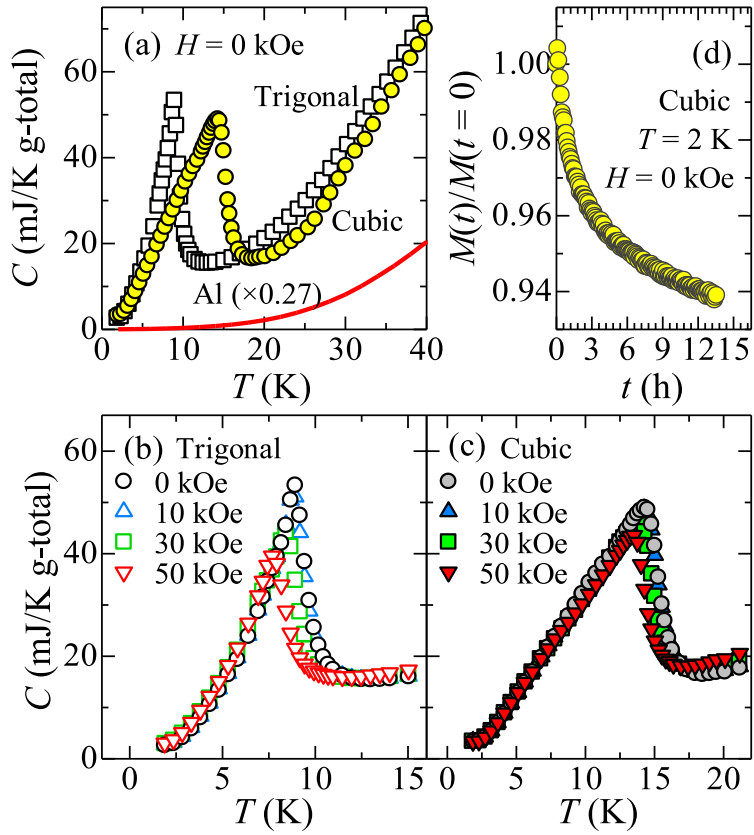}
\caption{(Color Online) (a) Temperature dependence of the specific heat $C$ at 0 kOe
for the trigonal HoAl$_{3}$ and the cubic HoAl$_{3}$.
The solid line represents $C$ of aluminum scaled by the weight fraction of 0.27.
(b)-(c) $C$ under several magnetic fields
in the trigonal HoAl$_{3}$ (b) and the cubic HoAl$_{3}$ (c).
(d) Magnetic relaxation curve at 2 K at 0 kOe in the cubic HoAl$_{3}$.}
\label{fig:HC&Mt}
\end{figure}

Figure \ref{fig:HC&Mt}(a) shows the temperature dependence of $C$ at 0 kOe
for the trigonal HoAl$_{3}$ and the cubic HoAl$_{3}$.
The specific heat data for pure Al, which was taken by ourselves, is also represented
by scaling with the weight fraction of 0.27.
The trigonal HoAl$_{3}$ exhibits a sharp specific heat peak near 10 K,
being associated with the AFM transition at $T^{\rm tri}_{\rm N}$.
By applying the magnetic field, the peak is slightly shifted to lower temperatures,
as shown in Fig. \ref{fig:HC&Mt}(b).
This corresponds to the decrease in cusp temperature of $M$/$H$.
The $C$ of the cubic HoAl$_{3}$ at 0 kOe has a similar peak near 15 K.
Moreover, as shown in Fig. \ref{fig:HC&Mt}(c),
a slight peak shift with increasing $H$ is also observed.
From the similarity between these two specific heat peaks,
we conclude that the cubic HoAl$_{3}$ undergoes an AFM transition
at the N$\acute{\rm e}$el temperature $T^{\rm cub}_{\rm N} =$ 15 K.
The large specific heat peaks near $T_{\rm N}$ guarantees that
the HoAl$_{3}$ phase is the main phase of each sample
used for the physical property measurements.
As expected above, 27 wt.\% of Al phase has a significant contribution
to the specific heat at above 20 K,
while it can be negligible at even lower temperatures.
The contribution from the HoAl$_{2}$ phase is not observed in both the samples.

For the cubic HoAl$_{3}$, no specific heat peak is found at 11 K
despite the clear cusp of $M$/$H$ is observed at 0.1 kOe at this temperature.
This discrepancy between $C$ and $M$/$H$ strongly suggests
the occurrence of a magnetic glassy state at 11 K in this compound
\cite{Mydosh-RPP-2015}.
To get more insight into the ground state,
we performed the magnetization relaxation measurement.
Here, the sample was first cooled down to 2 K at 0 kOe
and magnetized by the magnetic field of 50 kOe.
The magnetic field was then switched off after waiting for 5 minutes,
and finally a magnetic relaxation curve was measured at 0 kOe.
As can be seen from Fig. \ref{fig:HC&Mt}(d),
the magnetization continues to decrease over a long period of time.
This result also supports the presence of a magnetic glassy state.
Therefore, it is revealed that the magnetic state of the cubic HoAl$_{3}$ changes
from a long-range AFM ordered state to a magnetic glassy state
at a spin freezing temperature $T_{\rm f}$ of 11 K.
The nature and the origin of the magnetic glassy state are to be further verified
in a single-phase polycrystalline sample or a single crystal.

\subsection{Magnetocaloric properties}
\begin{figure}[t]
\centering
\includegraphics[width=70mm]{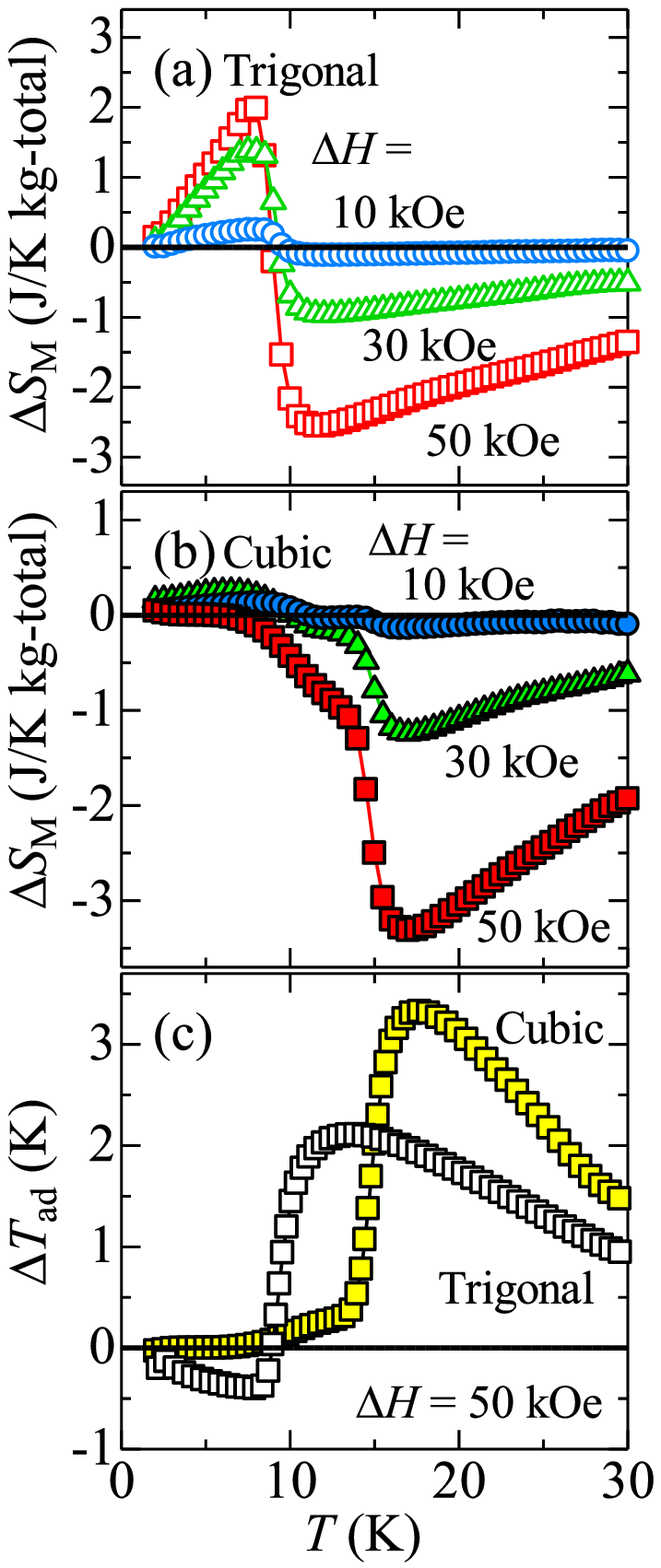}
\caption{(Color Online) (a)-(b) Magnetic entropy change $\Delta S_{\rm M}$ of
the trigonal HoAl$_{3}$ (a) and the cubic HoAl$_{3}$ (b)
for the magnetic field change $\Delta H =$ 10, 30, and 50 kOe.
(c) Adiabatic temperature change $\Delta T_{\rm ad}$ of the samples for $\Delta H =$ 50 kOe.}
\label{fig:MCE}
\end{figure}

Now let us evaluate the magnetocaloric properties of the two compounds.
In this study, we calculate the magnetic entropy change $\Delta S_{\rm M}$ from
the specific heat data by using the expression,
\begin{equation}
\Delta S_{\rm M} (\Delta H, T) = S(H, T) - S(0, T)
= \int_{0}^{T}\frac{C(H, T)}{T}dT - \int_{0}^{T}\frac{C(0, T)}{T}dT.
\end{equation}
Figure \ref{fig:MCE}(a) shows the temperature dependence of $\Delta S_{\rm M}$
of the trigonal HoAl$_{3}$ for $\Delta H =$ 10, 30, and 50 kOe.
At the temperature above $T^{\rm tri}_{\rm N} =$ 9.8 K,
negative $\Delta S_{\rm M}$ increases in magnitude with decreasing temperature
and is maximized at just above $T^{\rm tri}_{\rm N}$.
The maximum value is about $-$2.6 J kg$^{-1}$ K$^{-1}$ for $\Delta H =$ 50 kOe.
This is a conventional MCE expected in the paramagnetic region,
which reflects the release of the magnetic entropy
as the paramagnetic moments are aligned by a magnetic field.
As the temperature further drops,
$\Delta S_{\rm M}$ rapidly changes the sign to positive,
peaks near $T^{\rm tri}_{\rm N}$ in each magnetic field, and then goes to zero.
The maximum value is about $+$2.0 J kg$^{-1}$ K$^{-1}$ for $\Delta H =$ 50 kOe.
The positive $\Delta S_{\rm M}$ means
the restoration of magnetic entropy by applying the magnetic field.
This phenomenon---inverse MCE---is generally expected in AFM materials
\cite{Schelleng-JAP-1963,Biswas-JAP-2013},
being associated with antiparallel disorder of magnetic sublattices
\cite{Ranke-JPCM-2009}.

The $\Delta S_{\rm M}$-$T$ curves of the cubic HoAl$_{3}$ are shown
in Figure \ref{fig:MCE}(b).
The conventional MCE is observed in the paramagnetic region
at above $T^{\rm cub}_{\rm N} =$ 15 K.
The maximum negative value of $\Delta S_{\rm M}$ is about
$-$3.3 J kg$^{-1}$ K$^{-1}$ for $\Delta H =$ 50 kOe.
Although $\Delta S_{\rm M}$ drops towards
a positive value with further decreasing temperature,
the fall stops immediately
and the significant positive $\Delta S_{\rm M}$ peak is never found.
This is in quite contrary to the expectation that
like the trigonal HoAl$_{3}$,
an inverse MCE appears at below $T^{\rm cub}_{\rm N}$
in response to the disruption of a long-range AFM order by the magnetic field.
Instead, a hump appears around $T_{\rm f} =$ 11 K
and contributes to $\Delta S_{\rm M}$ as a negative component,
of which the magnitude increases with increasing $\Delta H$.
The replacement of the expected positive $\Delta S_{\rm M}$ peak by the hump also supports
that a long-range AFM order no longer exists in the cubic HoAl$_{3}$ at low temperatures,
but rather other magnetic state emerges.
Such succesive change in magnetic state leads to
the complicated behavior of $\Delta S_{\rm M}$ at below $T_{\rm f}$:
$\Delta S_{\rm M}$ takes small positive values
for $\Delta H =$ 10 kOe and 30 kOe,
while it shifts to negative with further increasing $\Delta H$.

Here let us evaluate another important magnetocaloric parameter,
\textit{i.e.}, the adiabatic temperature change $\Delta T_{\rm ad}$,
which can also be calculated from the $S$-$T$ curves as follows:
\begin{equation}
\Delta T_{\rm ad}(\Delta H, T) = T(H, S) - T(0, S).
\end{equation}
Figure \ref{fig:MCE}(c) shows the $\Delta T_{\rm ad}$ of each sample
as a function of temperature for $\Delta H =$ 50 kOe.
For both samples, the positive $\Delta T_{\rm ad}$, representing the conventional MCE, is
observed in the paramagnetic region,
peaking a little above the $T_{\rm N}$ of each sample.
The maximum value of positive $\Delta T_{\rm ad}$ is about 2.1 K for the trigonal HoAl$_{3}$
and 3.3 K for the cubic HoAl$_{3}$, respectively.
Below $T^{\rm cub}_{\rm N}$, a negative $\Delta T_{\rm ad}$ appears in the cubic HoAl$_{3}$,
indicating a decrease in temperature by adiabatic magnetization.
As a consequence of the small positive $\Delta S_{\rm M}$,
however, the maximum value is only about $-$0.4 K.
For the trigonal HoAl$_{3}$, no negative $\Delta T_{\rm ad}$ is found below $T^{\rm tri}_{\rm N}$,
but a small positive component is observed around 11 K,
corresponding to the hump of $\Delta S_{\rm M}$.
Note that the contribution of the secondary phase (Al phase) to the specific heat,
in addition to the weight fraction, may lead
a decrease in the value of $\Delta T_{\rm ad}$ observed for the samples
compared to the real value for the primary phase (HoAl$_{3}$ phase),
as has been pointed out in the previous work
\cite{TDY-JMMM-2020}.

It has been proposed that potential applications of inverse MCE materials would be
the cooling by adiabatic magnetization
or the use as a sink for the heat generated by a conventional MCE material
magnetized prior to cooling by adiabatic demagnetization
\cite{Joenk-JAP-1963,Krenke-NatMat-2005}.
Nevertheless, compared to conventional MCEs,
there are only a few reports on large inverse MCEs in AFM materials,
especially in the low-temperature region
\cite{Reis-JPCM-2010,Zuo-JAC-2013,Zheng-JMMM-2017}.
Not only that, but rather a conventional MCE is often observed for a large $\Delta H$
as a result of a field-induced metamagnetic transition
from an AFM state to a (forced) ferromagnetic state
\cite{Reis-JPCM-2010,Zuo-JAC-2013,Zheng-JMMM-2017,Chen-SSC-2010,Zhang-Intermet-2018}.
Owing to the robustness of the AFM ordering to the magnetic field,
the positive $\Delta S_{\rm M}$ is observed in the trigonal HoAl$_{3}$
even for $\Delta H =$ 50 kOe.
At the same time, however,
the inverse MCE is unfortunately not so large due to its robustness.
As for the cubic HoAl$_{3}$, the situation is somewhat complicated,
but at least it seems unlikely that
this compound is a ferromagnet with a large conventional MCE.

\section{Conclusions}
The magnetic, thermal, and magnetocaloric properties have been investigated
for two types of HoAl$_{3}$ with different crystallographic phases.
It has been found that the crystal structure of HoAl$_{3}$ is transformed
by the rapid-solidification process,
from the room-temperature equilibrium trigonal phase
to the high-pressure cubic phase.
The trigonal phase HoAl$_{3}$ is antiferromagnetically ordered
at below $T^{\rm tri}_{\rm N} =$ 9.8 K.
The AFM ordering is so robust against the magnetic field
that the inverse MCE is observed
even under a large magnetic field change of 0--50 kOe.
The cubic phase HoAl$_{3}$ undergoes the AFM transition at $T^{\rm cub}_{\rm N} =$ 15 K,
whereas the magnetic state immediately changes to a magnetic glassy state
below the spin freezing temperature $T_{\rm f} =$ 11 K.
As a consequence, the MCE exhibits complicated temperature and magnetic field dependence.
Our findings demonstrate that HoAl$_{3}$ compound cannot be
a promising magnetocaloric material,
unlike other binary holmium aluminides.

\section*{Acknowledgements}
This work was supported by JST-Mirai Program Grant Number JPMJMI18A3, Japan.

\newpage

\newpage

\begin{thebibliography}{99}
\bibitem{Pecharsky-PRL-1997}
	V. K. Pecharsky, K. A. Gschneidner, Jr.,
	Phys. Rev. Lett. 78 (1997) 4494-4497.

\bibitem{Gschneidner-RPP-2005}
	K. A. Gschneidner, Jr., V. K. Pecharsky, A. O. Tsokol,
	Rep. Prog. Phys. 68 (2005) 1479-1539.

\bibitem{Kitanovski-Springer-2015}
	A. Kitanovski, J. Tusek, U. Tomc, U. Plaznik, M. Ozbolt, A. Poredos,
	Springer International Publishing 2015.

\bibitem{Lyubina-JPD-2017}
	J. Lyubina,
	J. Phys. D: Appl. Phys. 50 (2017) 053002.

\bibitem{Li-SSC-2014}
	D. X. Li, T. Yamaura, S. Nimori, Y. Homma, F. Honda, Y. Haga, D. Aoki,
	Solid State Commun. 193 (2014) 6-10.

\bibitem{Matsumoto-JMMM-2017}
	K. T. Matsumoto, K. Hiraoka,
	J. Magn. Magn. Mater 423 (2017) 318-320.

\bibitem{Zhang-PhysicaB-2019}
	H. Zhang, R. Gimaev, B. Kovalev, K. Kamilov, V. Zverev, A. Tishin,
	Physica B: Condens. Matter 558 (2019) 65-73.

\bibitem{Zhang-JAC-2019}
	Y. Zhang,
	J. Alloys Compd. 787 (2019) 1173-1186.

\bibitem{Pedro-NPG-2020}
	P. B. Castro, K. Terashima, T. D. Yamamoto, Z. Hou, S. Iwasaki, R. Matsumoto,
	S. Adachi, Y. Saito, P. Song, H. Takeya, Y. Takano,
	NPG Asia Mater. 12 (2020) 35.

\bibitem{Li-JAC-2020}
	L. Li, M. Yan,
	J. Alloys Compd. 823 (2020) 153810.

\bibitem{Omote-Cryogenics-2019}
	H. Omote, S. Watanabe, K. Matsumoto, I. Gilmutdinov, A. Kiiamov, D. Tayuskii,
	Cryogenics 101 (2019) 58-62.

\bibitem{Li-ActMat-2020}
	L. Li, P. Xu, S. Ye, Y. Li, G. Liu, D. Hou, M. Yan,
	Acta Mater. 194 (2020) 354-365.

\bibitem{Wu-CerInt-2021}
	B. Wu, Y. Zhang, D. Guo, J. Wang, Z. Ren,
	Cer. Int. 47 (2021) 6290-6297.

\bibitem{Xu-MatTod-2021}
	P. Xu, Z. Ma, P. Wang, H. Wang, L. Li,
	Mat. Today Phys. 20 (2021) 100470.

\bibitem{Zhang-ActMat-2022}
	Y. Zhang, Y. Tian, Z. Zhang, Y. Jia, B. Zhang, M. Jiang, J. Wang, Z. Ren,
	Acta Mater. 226 (2022) 117669.


\bibitem{Hahimoto-ACEM-1986}
	T. Hashimoto, K. Matsumoto, T. Kurihara, T. Numazawa, A. Tomokiyo, H. Hayama,
	T. Goto, S. Todo, M. Shahari,
	Adv. Cryog. Eng. Mater. 32 (1986) 279-286.

\bibitem{Yang-JAC-2014}
	L. H. Yang, H. Zhang, F. X. Hu, J. R. Sun, L. Q. Pan, B. G. Shen,
	J. Alloys Compd. 596 (2014) 58-62.

\bibitem{Zhang-SSC-2012}
	H. Zhang, Z. Y. Xu, X. Q. Zheng, J. Shen, F. X. Hua, J. R. Sun, B. G. Shen,
	Solid State Commun. 152 (2012) 1127-1130.

\bibitem{Bhattacharyya-JPCM-2010}
	A. Bhattacharyya, S. Giri, S. Majumdar,
	J. Phys.: Condens. Matter 22 (2010) 316003.

\bibitem{Vucht-JLCM-1965}
	J. H. N. Van Vucht, K. H. J. Buschow,
	J. Less-Common Metals 10 (1965) 98-107.


\bibitem{Cannon-JLCM-1975}
	J. F. Cannon, H. T. Hall,
	J. Less-Common Metals 40 (1975) 313-328.

\bibitem{Buschow-HFM-1980}
	K. H. J. Buschow,
	Rare earth compounds,
	in: E. P. Wohlfarth (Ed.),
	Handbook of Ferromagnetic Materials, vol. 1, Elsevier, 1980, pp. 297-414.

\bibitem{Sahashi-IEEETrans-1987}
	M. Sahashi, H. Niu, Y. Tohkai, K. Inomata, T. Hashimoto,
	T. Kuzuhara, A. Tomokiyo, H. Yamaya,
	IEEE Trans. Magn. 23 (1987) 2853-2855.

\bibitem{Momma-VESTA-2011}
	K. Momma, F. Izumi,
	J. Appl. Crystallogr. 44 (2011) 1272.

\bibitem{Xu-APL-1991}
	Y. Xu, Z. Altounian, W. B. Muir,
	Appl. Phys. Lett. 58 (1991) 125-127.

\bibitem{Meyer-JLCM-1966}
	A. Meyer,
	J. Less-Common Metals 10 (1966) 121-129.

\bibitem{Pedro-2022}
	P. B. Castro, K. Terashima, M. G. E. Echevarria, H. Takeya, Y. Takano,
	Adv. Theory and Simul (2021): 2100588.

\bibitem{Hu-APL-2008}
	W. J. Hu, J. Du, B. Li, Q. Zhang, Z. D. Zhang,
	Appl. Phys. Lett. 92 (2008) 192505.

\bibitem{Khan-JAP-2013}
	M. Khan, D. Paudyal, K. A. Gschneidner, Jr., V. K. Pecharsky,
	J. Appl. Phys. 113 (2013) 17E106.


\bibitem{Mydosh-RPP-2015}
	J. A. Mydosh,
	Rep. Prog. Phys. 78 (2015) 052501.

\bibitem{Schelleng-JAP-1963}
	J. H. Schelleng, S. A. Friedberg,
	J. Appl. Phys. 34 (1963) 1087-1089.

\bibitem{Biswas-JAP-2013}
	A. Biswas, S. Chandra, T. Samanta, M. H. Phan, I. das, H. Srikanth,
	J. Appl. Phys. 113 (2013) 17A902.


\bibitem{Ranke-JPCM-2009}
	P. J. Von Ranke, N. A. de Oliveira, B. P. Alho, E. J. R. Plaza,
	V. S. R. de Sousa, L. Caron, M. S. Reis,
	J. Phys.: Condens. Matter 21 (2009) 056004.

\bibitem{TDY-JMMM-2020}
	T. D. Yamamoto, H. Takeya, K. Terashima, S. Iwasaki, P. B. Castro,
	T. Numazawa, Y. Takano,
	J. Magn. Magn. Mater. 513 (2020) 167207.

\bibitem{Joenk-JAP-1963}
	R. J. Joenk,
	J. Appl. Phys. 34 (1963) 1097-1098.
	
\bibitem{Krenke-NatMat-2005}
	T. Krenke, E. Duman, M. Acet, E. F. Wassermann, X. Moya, L. Ma$\tilde{\rm n}$osa, A. Planes,
	Nat. Mat. 4 (2005) 450-454.

\bibitem{Reis-JPCM-2010}
	R. D. does Reis, L. M. da Silva, A. O. dos Santos, A. M. N. Medina,
	L. P. Cardoso, F. G. Gandra,
	J. Phys.: Condens. Matter 22 (2010) 486002.

\bibitem{Zuo-JAC-2013}
	W. Zuo, F. Hu, J. Sun, B. Shen,
	J. Alloys Compd. 575 (2013) 162-167.

\bibitem{Zheng-JMMM-2017}
	X. Q. Zheng, Z. Y. Xu, B. Zhang, F. X. Hu, B. G. Shen,
	J. Magn. Magn. Mater. 421 (2017) 448-452.

\bibitem{Chen-SSC-2010}
	J. Chen, B. G. Shen, Q. Y. Dong, J. R. Sun,
	Solid State Commun. 150 (2010) 1429-1431.

\bibitem{Zhang-Intermet-2018}
	Y. Zhang, Y. Yang, C. Hou, D. Guo, X. Li, Z. Ren, G. Wilde,
	Intermetallics 94 (2018) 17-21.

	
\end{thebibliography}
\end{document}